\documentclass{epl}

\title{Phase noise due to vibrations in Mach-Zehnder atom interferometers}
\shorttitle{Phase noise in atom interferometers}
\author{A. Miffre\inst{1,2} \and M. Jacquey\inst{1} \and M. B\"uchner\inst{1}
\and G. Tr\'enec\inst{1} \and J. Vigu\'e\inst{1}
\thanks{E-mail:\email{jacques.vigue@irsamc.ups-tlse.fr}}}

\institute{ \inst{1} Laboratoire Collisions Agr\'egats
R\'eactivit\'e -IRSAMC
\\Universit\'e Paul Sabatier and CNRS UMR 5589
\\118, Route de Narbonne 31062 Toulouse Cedex, France\\
\inst{2} PIIM, Universit\'e de Provence and CNRS UMR 6633,
\\ Centre de saint J\'er\^ome case C21, 13397 Marseille cedex 20,
France}

\date{\today}
\begin{document}

\maketitle

\begin{abstract}

Atom interferometers are very sensitive to accelerations and
rotations. This property, which has some very interesting
applications, induces a deleterious phase noise due to the seismic
noise of the laboratory and this phase noise is sufficiently large
to reduce the fringe visibility in many experiments. We develop a
model calculation of this phase noise in the case of Mach-Zehnder
atom interferometers and we apply this model to our thermal
lithium interferometer. We are able to explain the observed phase
noise which has been detected through the rapid dependence of the
fringe visibility with the diffraction order. We think that the
dynamical model developed in the present paper should be very
useful to reduce the vibration induced phase noise in atom
interferometers, making many new experiments feasible.
\end{abstract}


\section{Introduction}

Atom interferometers are very sensitive to inertial effects
\cite{anandan77,clauser88} and this property was used to build
accelerometers
\cite{kasevich91,kasevich92,cahn97,peters99,peters01,snadden98,mcguirk02,tino02}
and gyrometers
\cite{riehle91,lenef97,gustavson97,gustavson00,leduc04}. Because
of this large sensitivity, a high mechanical stability of the
experiment is required and several experiments
\cite{keith91,giltner95b,peters99,peters01} used an active control
of the interferometer vibrations.

In the present letter, we study the phase noise induced by
mechanical vibrations in three-gratings Mach-Zehnder thermal atom
interferometers and we show that the rapid decrease of the fringe
visibility with the diffraction order is largely due to this phase
noise. Vibrations displace and bend the rail which holds the three
diffraction gratings and we have developed a model based on
elasticity theory to describe this dynamics. We are thus able to
understand the contributions of various frequencies and to make a
detailed evaluation of this phase noise in the case of our setup:
the result is in good agreement with the phase noise value deduced
from fringe visibility measurements. We have built a very stiff
rail for our atom interferometer and this arrangement has revealed
very efficient: the remaining phase noise is dominated by
rotations of the rail, which should be reduced by a better rail
suspension.

\section{Fringe visibility as a test of phase noise in atom
interferometers}

A phase noise $ \Phi(t) $ has the effect of reducing the fringe
visibility $  {\mathcal{V}} =\left(I_{max} -
I_{min}\right)/\left(I_{max} + I_{min}\right)$. Assuming a
Gaussian distribution of $\Phi$, the visibility is given by
\cite{schmiedmayer97,delhuille02}

\begin{equation}
\label{n1} {\mathcal{V}}   = {\mathcal{V}}_{max} \exp\left[-
\left<\Phi^2 \right>/2 \right]
\end{equation}

\noindent When the phase noise is due to inertial effects, we
prove below that it is proportional to the diffraction order $p$,
$\Phi_p = p \Phi_1$, where $\Phi_p$ corresponds to the order $p$.
Equation (\ref{n1}) predicts a Gaussian dependence of the fringe
visibility ${\mathcal{V}}$ with the diffraction order $p$
\cite{delhuille02}:

\begin{equation}
\label{n2}{\mathcal{V}}  = {\mathcal{V}}_{max} \exp\left[-p^2
\left<\Phi_1^2 \right>  /2\right]
\end{equation}

\begin{figure}
\onefigure[width=8cm]{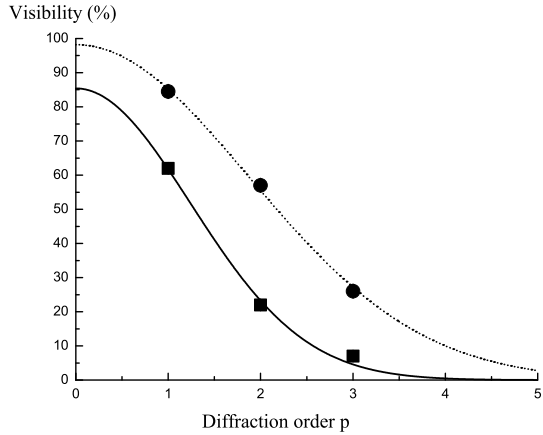} \caption{Fringe visibility as a
function of the diffraction order $p$. Our measurements (round
dots) are fitted by equation \ref{n2} with ${\mathcal{V}}_{max}
=98\pm 1$ \% and $\left<\Phi_1^2 \right> = 0.286 \pm 0.008$. The
data points of Giltner and Siu Au Lee (squares) are also fitted by
equation \ref{n2} with ${\mathcal{V}}_{max} =85\pm 2$ \% and
$\left<\Phi_1^2 \right> = 0.650 \pm 0.074$.} \label{fig1}
\end{figure}

\noindent Only two atom interferometers have been operated with
several diffraction orders, by Siu Au Lee and co-workers
\cite{giltner95b,giltner96} in 1995 and more recently by our group
\cite{miffre05}. The observed fringe visibility is plotted as a
function of the diffraction order $p$ in figure \ref{fig1}. A
Gaussian fit, following equation (\ref{n2}), represents very well
the data in both cases and the quality of this fit suggests the
importance of a phase noise from inertial origin.

\section{Inertial sensitivity of Mach-Zehnder atom interferometers}

\begin{figure}
\onefigure[width=8cm]{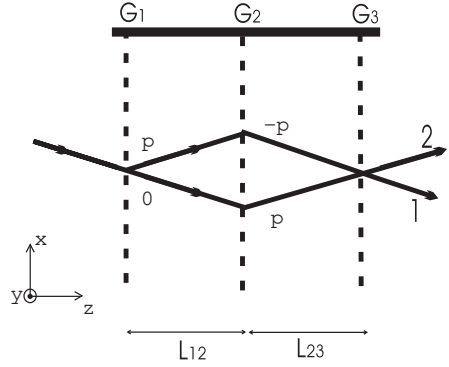}\caption{Schematic drawing of a
three grating Mach-Zehnder atom interferometer, in the Bragg
diffraction geometry. A collimated atomic beam is successively
diffracted by three gratings $G_1$, $G_2$ and $G_3$. The
diffraction orders corresponding to grating $G_1$ and $G_2$ are
indicated on the two atomic beams. Two exit beams, labelled $1$
and $2$, carry complementary signals. The $x$, $y$, $z$ axis are
defined.} \label{MZschematic}
\end{figure}

We consider a three-grating Mach-Zehnder interferometer
represented in figure \ref{MZschematic}. The inertial sensitivity
of this type of interferometer is due to the existence of the
diffraction phase which depends on the grating positions. The
resulting phase of the interference signal is given by
\cite{schmiedmayer97,delhuille02}:

\begin{equation}
\label{n3} \Phi_p = p k_G \left[2 x_2(t_2) - x_1(t_1) -x_3(t_3)
\right]
\end{equation}

\noindent Here $k_G = 2 \pi/a $ is the grating wavevector ($a$
being the grating period); $p$ is the diffraction order and
$x_j(t_j)$ the $x$-coordinate of a reference point of grating
$G_j$ at time $t_j$ when it is crossed by the atomic wavepacket.
Equation (\ref{n3}) can be simplified by introducing the atom time
of flight $T=L_{12}/u$ from one grating to the next, with $L_{12}
= L_{23}$ and $u$ being the atom velocity (we will neglect its
dispersion throughout the present paper). We can express the time
$t_j$ as a function of $t_2$, which will be noted $t$, and $T$:
$t_1 = t-T$ and $t_3 = t+T$. Then, if we expand $\Phi$ in powers
of $T$ up second order, we get:

\begin{equation}
\label{n4} \Phi =  p k_G\left[\delta(t) - \left[ v_{3x}(t) -
v_{1x}(t) \right] T
 - \frac{\left[ a_{1x}(t) + a_{3x}(t) \right] T^2}{2}\right]
 \end{equation}

\noindent Here $\delta(t) =2 x_2(t) - x_1(t) -x_3(t)$ while
$v_{jx}(t)$ and $a_{jx}(t)$ are the $x$-components of the velocity
and acceleration of grating $G_j$ measured by reference to a
Galilean frame. In equation (\ref{n4}), the first term is due to
the instantaneous bending $\delta(t)$ of the rail, the second term
represents Sagnac effect, as the velocity difference is related to
the rail angular velocity, and the third term describes the
sensitivity to accelerations.

\section{Theoretical analysis of the rail dynamics}

We want to relate the positions $x_j(t_j)$ of the three gratings
to the mechanical properties of the rail and to its coupling to
the seismic noise. As the interferometer is sensitive only to the
grating $x$-coordinates, we use a 1D model to describe the rail
dynamics. This model is based on elasticity theory \cite{landau86}
in order to describe, in an unified way, the motion and the
deformations of the rail. In this model, the rail of length $2L$
along the $z$ direction can bend only in the $x$ direction. The
shape of its cross-section, assumed to be independent of the
$z$-coordinate, is characterized by its area $A= \int dx dy$ and
by the moment $I_y = \int x^2 dx dy$, the $x$-origin being taken
on the neutral line. The rail material has a density $\rho$ and a
Young's modulus $E$. The neutral line is described by a function
$X(z,t)$ which measures the position of this line with respect to
a Galilean frame  and which verifies \cite{landau86}:

\begin{equation}
\label{elasticity1} \rho A \frac{\partial^2 X}{\partial t^2} = -E
I_y \frac{\partial^4 X}{\partial z^4}
\end{equation}

\noindent The rail is coupled to the laboratory by forces and
torques exerted at $z= \epsilon L$ ($\epsilon = \pm$) by its
supports. The $x$-component of the force $F_{x\epsilon}$ and the
$y$-component of the torque are respectively related
\cite{landau86} to the third and second derivatives of $X(z,t)$
with respect to $z$. We assume that the torques vanish so that $
\partial^2 X/\partial z^2 = 0$ at $z= \epsilon L$ and that the
force is the sum of an elastic term proportional to the relative
displacement and a damping term proportional to the relative
velocity:
\begin{equation}
\label{elasticity3} F_{x\epsilon} = - \epsilon E I_y
\frac{\partial^3 X}{\partial z^3}(z= \epsilon L) = -
K\left[X(\epsilon L,t) -x_{\epsilon}(t)\right] - \mu
\frac{\partial\left[X(\epsilon L,t) -x_{\epsilon}(t)
\right]}{\partial t}
\end{equation}

\noindent  where $x_{\epsilon}(t) $ is the $x$-position of the
support at $z =\epsilon L$ and time $t$. We introduce the Fourier
transforms $X(z,\omega)$ and $x_{\epsilon}(\omega)$ of the
functions $X(z,t)$ and $x_{\epsilon}(t)$. The general solution of
equation (\ref{elasticity1}) is:

\begin{equation}
\label{solution1} X(z,\omega) = a \sin(\kappa z) + b\cos(\kappa z)
+ c \sinh(\kappa z) + d\cosh(\kappa z)
\end{equation}

\noindent $a$, $b$, $c$ and $d$ are the four $\omega$-dependent
components of $X(z,\omega)$. $c$ and $d$ are related to $a$ and
$b$, thanks to assumption of vanishing torques. Then $a$ and $b$
are linearly related to the source terms $x_{\epsilon}(\omega)$ by
equations (\ref{elasticity3}). $\omega$ and $\kappa$ are related
by:

\begin{equation}
\label{solution2} \rho A \omega^2 = E I_y \kappa^4
\end{equation}

When $\mu$ is small enough, these equations predict a series of
resonances. The first resonance, when $\omega= \omega_{osc}
=\sqrt{K/(\rho A L)}$, describes an in-phase oscillation of the
two ends of the rail while the second resonance, occurring when
$\omega = \omega_{osc}\sqrt{3}$, describes a rotational
oscillation of the rail around its center. Then, there is an
infinite series of bending resonances occurring for $\kappa_n$
verifying $\cos(2\kappa_n L) \cosh(2\kappa_n L) =1$ and $\omega_n$
related to the $\omega_0$, by $\omega_n =\omega_0
(\kappa_n/\kappa_0)^2$. In our model, the rail stiffness is
described by one parameter only, namely $\omega_0$.

\begin{equation}
\label{solution11} \omega_0 = 5.593 \sqrt{EI_y/(\rho A L^4)}
\end{equation}

\section{ The phase noise due to vibrations}

The Fourier component $\Phi_p(\omega)$ of the phase $\Phi_p$ given
by equation (\ref{n3}) can be expressed as a function of the
amplitudes $a$ and $b$. We assume that the grating reference
points are on the neutral line, at $z = \epsilon L_{12}$
($\epsilon=\pm$) and we get:
\begin{eqnarray}
\label{excitation1} \Phi_p(\omega)/p =  2 k_G  &  & \left[
b(\omega) \left(1 -\cos(\kappa L_{12})  + \left(1 -\cosh(\kappa
L_{12}) \right) \frac{\cos(\kappa L)}{\cosh(\kappa
L)}\right)\right. \nonumber \\ & & + \left. i a(\omega) \left(
\sin(\kappa L_{12}) + \sinh(\kappa L_{12}) \frac{\sin(\kappa
L)}{\sinh(\kappa L)} \right) \sin\left(\omega T \right) \right.
\nonumber \\ & & + \left. b(\omega) \left( \cos(\kappa L_{12})
+\cosh (\kappa L_{12}) \frac{\cos(\kappa L)}{\cosh(\kappa L)}
\right)  \left( 1 -\cos(\omega T)\right)\right]
\end{eqnarray}

\noindent where the different lines correspond to the bending, the
Sagnac and the acceleration terms in this order. This complicated
equation can be given a very simple form by making expansions in
powers of $(\omega T)$ and $\kappa L$ (assuming $L_{12} =L$ for
further simplification):

\begin{eqnarray} \label{excitation4}
\Phi_p(\omega)/p & \approx & k_G \times \left[ \left[x_{+}
(\omega) -x_{-}(\omega) \right] \frac{3i (\omega
T)}{\left(3-R\right)}
 \right. \nonumber
\\ & + & \left. \left[x_{+}(\omega)  + x_{-}(\omega) \right]
 \frac{ 13.0 (\omega/\omega_0)^2 + (\omega T)^2 }{2(1-R)} \right]
\end{eqnarray}

\noindent where $R = \omega^2/\left[\omega_{osc}^2
-i(\omega_{osc}\omega/Q_{osc})\right]$. Equation
(\ref{excitation4}) has a limited validity because of numerous
approximations but it gives a very clear view of the various
contributions. The first term, proportional to $\left[x_{+}
(\omega) -x_{-}(\omega)\right]$ and to the time of flight $T$,
describes the effect of the rotation of the rail excited by the
out of phase motion of its two ends. This term, which is
independent of the stiffness of the rail, is sensitive to the rail
suspension through the $(3-R)$ denominator. The second term is the
sum of the bending term, in $(\omega /\omega_0)^2 $, and the
acceleration term, in $(\omega T)^2 $. Both terms have the same
sensitivity to the suspension of the rail, being sensitive to the
first pendular resonance, when $R\approx 1$. The bending term is
small if the rail is very stiff, i.e. when $\omega_0$ is large.

\section{Application of the present analysis to our interferometer}

When we built our interferometer, we knew that D. Pritchard
\cite{keith91,schmiedmayer97} and Siu Au Lee
\cite{giltner95b,giltner96} had been obliged to reduce $\delta(t)$
in their atom interferometers by a servo-loop. Rather than using a
servo-loop, we decided to improve the grating stability by
building a very stiff rail. We use aluminium alloy for its large
$E/ \rho$ ratio and we made the largest possible rail in the $x$
direction to get a large $I_y/A$ ratio. Using equation
(\ref{solution11}), we estimate $\omega_0/2 \pi \approx 437$ Hz,
in reasonable agreement with our measurement, $\omega_0/(2\pi) =
460.4$ Hz, with a rather large Q-factor, $Q \approx 60$ (more
details in \cite{miffre06}). The suspension of the rail is very
simple, with rubber blocks made to support machine tools. From a
rough estimate of their force constant $K$ and the rail mass, the
first resonance is calculated to be at $\omega_{osc}/(2\pi)
\approx 20$ Hz.

Following previous works
\cite{gruber89,rasel95,keith91,schmiedmayer97,giltner95b}, the
instaneous bending $\delta(t)$ is conveniently measured by a
3-grating Mach-Zehnder optical interferometer attached to the
gratings of the atom interferometer. We have built such an
interferometer \cite{miffre02}, with $200$ lines/mm gratings from
Paton Hawksley \cite{paton} ($k_{g,opt} = 3.14\times10^5$
m$^{-1}$) and an helium-neon laser at $633$ nm. The phase
$\Phi_{opt}$ of the signal of such an optical interferometer is
also given by equation (\ref{n4}) simplified because the time of
flight $T$ for light is negligible: $\Phi_{opt} = p k_{g,opt}
\delta(t)$. In our experiment, the excitation of the rail by the
environment gives very small signals, from which we deduce an
upper limit of the bending $\sqrt{\left< \delta(t)^2\right>} <3$
nm.

\begin{figure}
\onefigure[width=8cm]{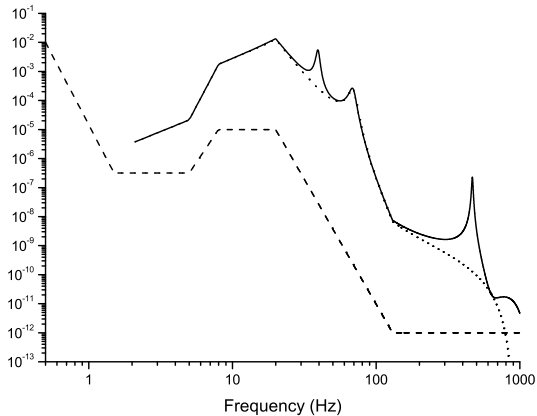} \caption{Calculated phase noise
spectra $\left|\Phi(\nu)/p\right|^2$ (full curve) and
$\left|\Phi_{Sagnac}(\nu)/p\right|^2$ (dotted curve), both in
rad$^2/$Hz as a function of the frequency $\nu$ in Hz. The
smoothed seismic noise spectrum $|x_{\epsilon}(\nu)|^2$ in
m$^2/$Hz used in the calculation is plotted (dashed curve) after
multiplication by $10^{10}$. The spectrum was recorded in the $\nu
=0.5-100$ Hz range and assumed to be constant when $10^2< \nu
<10^3$ Hz.} \label{noisespectra}
\end{figure}


To evaluate the phase noise, we need to know the seismic noise
spectrum. A spectrum was recorded on our setup well before the
operation of our interferometer and we use this measurement as a
good estimate of the seismic noise. We have replaced the recorded
spectrum with several peaks appearing in the $8-60$ Hz range by a
smooth spectrum just larger than the measured one. Most of the
peaks do not appear on a spectrum taken on the floor, because they
are due due to resonances of the structure supporting the vacuum
pipes and their exact frequency has probably changed because of
modifications of the experiment since the recording. The smoothed
noise spectrum $|x_{\epsilon}(\nu)|^2$ is plotted in figure
\ref{noisespectra}. This figure also plots the calculated phase
noise spectrum $\left|\Phi(\nu)/p\right|^2$, using equation
(\ref{excitation1}), and the Sagnac phase noise spectrum
$\left|\Phi_{Sagnac}(\nu)/p\right|^2$ deduced from equation
(\ref{excitation1}) by keeping only the term proportional to the
$a$ amplitude: clearly, Sagnac phase noise is dominant except near
the in-phase pendular oscillation and the first bending resonance.
The contribution of the in-phase pendular oscillation depends
strongly on its frequency and $Q$-factor. The bending resonance is
in a region where the excitation amplitude is very low, and, even
after amplification by the resonance $Q$-factor, the contribution
of the bending resonance to the total phase noise is fully
negligible.

In this calculation, we have not used our estimate of the first
pendular resonance $\omega_{osc}/(2\pi) \approx 20$ Hz, because
the predicted rms value of the bending $\sqrt{\left<
\delta(t)^2\right>}$ was considerably larger than the measured
upper limit. We have used $\omega_{osc}/(2\pi) = 40$ Hz, with
$Q_{osc}\approx 16$ and the measured $\omega_0$ value,
$\omega_0/(2\pi) = 460.4$ Hz and  $T= 5.7\times 10^{-4}$ s
($L_{12} = 0.605 m$ and $u= 1065$ m/s). We assume that the two
excitation terms $x_{\epsilon}(\nu)$ have the same spectrum but no
phase relation, so that we neglect the cross-term
$\left|x_{+}(\nu)x_{-}(\nu)\right|$. For very low frequencies up
to a few Hertz, we expect $x_{+}(\nu) \approx x_{-}(\nu)$ and the
associated correction would cancel the Sagnac term and this why we
have not extended the $\left|\Phi(\nu)/p\right|^2$ curves below
$2$ Hz. As soon as the frequency is larger than the lowest
resonance frequency of the structure supporting the vacuum
chambers (near $8$ Hz), the assumption that $x_{+}(\nu)$ and
$x_{-}(\nu)$ have no phase relation should be good.

By integrating the phase noise over the frequency from $2$ to
$10^3$ Hz, we get an estimate of the quadratic mean of the phase
noise:

\begin{equation}
\label{phasenoise1} \left<\Phi_p^2\right>  = 0. 16 p^2\mbox{
rad}^2
\end{equation}

\noindent  This estimate compares well with the value
$\left<\Phi_p^2 \right> = ( 0. 286 \pm 0.008) p^2$, deduced from
the fit of figure \ref{fig1}: we think that the agreement is
convincing, if ones considers the large uncertainty on several
parameters (seismic noise, frequency and $Q$ factors of the
pendular resonances), . This result is largely due to Sagnac phase
noise, as the same integration only on Sagnac phase noise gives
$\left<\Phi_{Sagnac}^2\right>  = 0. 13 p^2$ rad$^2$ and as shown
by equation (\ref{excitation4}), this phase noise can be reduced
only by a modification of the rail suspension.

\section{Conclusion}

 The  present paper has analyzed the phase noise
induced in a Mach-Zehnder atom interferometer by mechanical
vibrations (more details in \cite{miffre06}). Starting from the
well-known inertial sensitivity of atom interferometers, we have
developed a simple $1$D model describing the dynamics of the rail
holding the diffraction gratings. This model provides an unified
description of the low- and high-frequency dynamics, in which the
rail behaves respectively as a solid object and an elastic object.
In the low-frequency range, up to the frequency of the rotational
resonance of the suspension, the out-of-phase vibrations of the
two ends of the rail induce small rotations, which are converted
into phase noise by Sagnac effect, and this is the dominant cause
of inertial phase noise in our interferometer.

We think that the present analysis is important as it gives access
to a reduction of this phase noise in atom interferometers. A
better rail suspension should considerably reduce this phase
noise. Then, we would be able to observe atom interference effects
with an excellent fringe visibility, close to the fitted value
${\mathcal{V}}_{max} =98 \pm 1$ \% of figure \ref{fig1}, and we
would also be able to work either with diffraction orders $p \gg
1$ or with considerably slower atoms.

\acknowledgments

We have received the support of CNRS MIPPU, of ANR and of R\'egion
Midi Pyr\'en\'ees through a PACA-MIP network. We thank A. Souriau
and J-M. Fels for measuring the seismic noise in our laboratory.



\newpage

\end{document}